\documentclass[aps,prl,twocolumn,groupedaddress]{revtex4-1}

\usepackage{amsmath}           
\usepackage{graphicx}          
\usepackage{dcolumn}           

\begin{document}               

\title{Electron Quantization in Broken Atomic Wires}       

\author{Eui Hwan Do}		    
\author{Han Woong Yeom}
\email{yeom@postech.ac.kr}	    
\affiliation{Center for Artificial Low Dimensional Electronic Systems, Institute for Basic Science (IBS), 77 Cheongam-Ro, Pohang 790-784, Korea}
\affiliation{Department of Physics, Pohang University of Science and Technology (POSTECH), Pohang 790-784, Republic of Korea} 

\date{\today}                  

\begin{abstract}				
We demonstrate using scanning tunneling microscopy and spectroscopy the electron quantization within metallic Au atomic wires self-assembled on a Si(111) surface and segmented by adatom impurities. The local electronic states of wire segments with a length up to 10 nm are investigated as terminated by two neighboring Si adatoms. One dimensional (1D) quantum well states are well resolved by their spatial distributions and the inverse-length-square dependence in their energies. The quantization also results in the quantum oscillation of the conductance at the Fermi level. These results deny the dopant role of the adatoms assumed for a long time but indicate their strong scattering nature. The present approach provides a new and convenient platform to investigate 1D quantum phenomena with atomic precision. 
\end{abstract}

\maketitle			        

One-dimensional (1D) materials systems have been investigated extensively because both of the intriguing and exotic nature of their electrons and of the interest in nano or atomic scale devices. Characteristic 1D electronic properties of fundamental interest include the Tomonaga-Luttinger liquid \cite{ishii2003direct,Auslaender01022002}, the charge-density wave \cite{PhysRevLett.82.4898,zeng2008charge}, the unconventional superconductivity \cite{PhysRevLett.53.1179}, and more recently, Majorana Fermions \cite{das2012zero,Nadj-Perge31102014}. However, 1D materials systems are notorious in their intrinsic susceptibility to extrinsic perturbations such as defects and impurities~\cite{voit1995one,RevModPhys.84.1253,PhysRevB.90.165410}. In microscopic points of view, defects or impurities can have various different actions; local or global doping, local lattice distortions, weak or strong electron scatterings and so on. Nevertheless, the direct atomic scale investigation on those actions has been far from being sufficient.  

The self-assembled atomic wire structure of the Si(111)5$\times$2-Au surface is an excellent candidate system for a systematic study on the interactions of impurities with atomic scale precision. This system has a well-ordered Au-Si atomic wire array with a well-defined 1D metallic band \cite{PhysRevLett.100.126801}. Its atomic structure has been debated for a long time but very recently determined conclusively \cite{PhysRevLett.113.086101,PhysRevLett.113.165501}. This surface is intrinsically endowed with Si adatom impurities ejected from the Si substrate \cite{PhysRevB.67.205412,Kirakosian2003928}, whose density can be controlled globally \cite{PhysRevLett.100.126801} or in the atom-by-atom fashion \cite{bennewitz2002atomic}. The electronic band structure is systematically tuned by the adatom density from a metal to a gapped insulator globally \cite{PhysRevLett.100.126801} or locally \cite{PhysRevLett.92.096801}. This controllability opens up an unprecedented possibility for a systematic study of impurity effects in an atomic scale 1D metallic system \cite{PhysRevLett.100.146103}. 

For quite a long time, a Si adatom on this model 1D system was believed to act as a charged impurity \cite{PhysRevLett.91.206101} and a recent scanning tunneling microscopy and spectroscopy (STM/S) work claimed the observation of the local `confined' doping effect of individual adatoms \cite{PhysRevLett.109.066801}. Within this picture, an impurity atom act only to locally shift the energy of the 1D metallic state while the electron scattering and the Friedel oscillation \cite{Ono2009469} by the impurity are totally neglected. In this work, we focus on the electron scattering by a Si adatom. The electron scattering by a single impurity would induce 1D standing waves and that by two neighboring impurities could create the 1D quantum confinement. The latter is fully analogous to the vertical quantization in metallic ultrathin films \cite{PhysRevB.27.1991,PhysRevLett.80.5381,chiang2000photoemission,PhysRevLett.93.026802,Guo10122004,Nilius13092002,PhysRevLett.92.056803,PhysRevLett.103.096104,mocking2013electronically}, where quantum well states (QWS's) were spectroscopically observed and the density-of-states (DOS) oscillations at Fermi level was shown to drive thickness-dependent quantum oscillations of various physical and chemical properties \cite{PhysRevB.27.1991,PhysRevLett.80.5381,chiang2000photoemission,PhysRevLett.93.026802,Guo10122004}. In atomic scale 1D systems, most of previous works studied QWS's of atom-manipulated atomic chain systems on metallic surfaces \cite{Nilius13092002,PhysRevLett.92.056803,PhysRevLett.103.096104}. These works have difficulties in decoupling the electronic states at Fermi level from those of metallic substrates \cite{Nilius13092002,PhysRevLett.92.056803} and in extending the chain length \cite{mocking2013electronically}. Due partly to such difficulties, the length-dependent quantum oscillation has been rarely demonstrated in atomic chain systems.

In this Letter, we show that the Si adatoms on Si(111)5$\times2$-Au atomic wires act as strong scatterers to confine and quantize 1D electrons. The wire segments with a length between 1.5 and 10.5 nm as defined by two neighboring adatoms were investigated systematically by STM/S. The 1D QWS's are well resolved in both filled and empty states, whose spatial distributions and energy dependence on the wire length are fully consistent with the 1D quantum box model and the band structure calculated. This work clearly denies the dopant role of a Si adatom. The electron quantization is shown to give rise to a 4$a_{0}$ [$a_{0}$ = 0.384 nm, the Si(111) lattice constant] quantum oscillation of the zero bias conductance. This work introduces a new platform for the quantitative exploitation of various quantum properties of atomic scale 1D systems.

Experiments were performed using a commercial low-temperature STM (Unisoku, Japan) at 78 K. The clean Si(111)7$\times$7 surface was prepared by repeated flash heating to 1520 K. Gold was evaporated onto the clean Si(111)7$\times$7 surface held at 932 K. At the Au coverage of 0.6$-$0.7 monolayer, well-ordered 5$\times$2-Au surfaces were reproducibly formed \cite{PhysRevB.79.155301,PhysRevB.89.035416}. For STS measurements current-voltage ($I$/$V$) curves were recorded with the feedback loop switched off at a set point of 0.3 nA and -1.0 V.


\begin{figure}[!tb] 
\centering
\includegraphics[width=8.6cm]{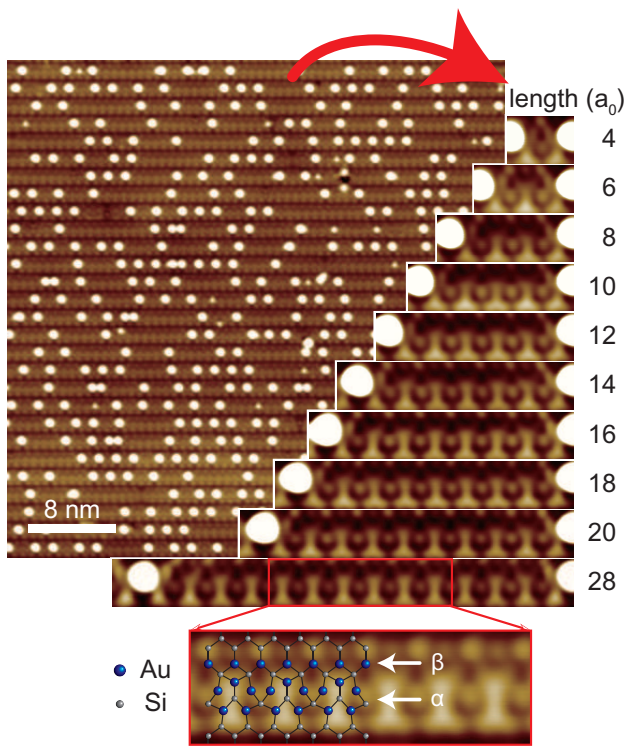}
\caption{\label{f1} Topographic STM image (45.3$\times$45.3 nm$^2$) of Si(111)5$\times$2-Au at 78.2 K and close-ups for wire segments between two Si adatoms of varying separation. The images are taken at -1.5 (large image) and -1.0 V (wire segments) sample bias. Schematics of the most recent structural model \cite{PhysRevLett.113.086101} is shown for the central part of the wire segment in the bottom (blue and silver spheres indicate Au and Si atoms, respectively).}
\end{figure}

Figure \ref{f1} shows STM images of Si(111)5$\times$2-Au surface. Well-ordered atomic wires with a 5$\times$2 unit cell are observed, which are decorated with Si adatoms appearing as bright protrusions. The atomic wires are segmented by Si adatoms, whose lengths (distance between two neighboring adatoms) are preferentially even multiples of $a_{0}$~\cite{PhysRevB.67.205412}. This is due to the atomic structure of the wire with a 2$a_{0}$ unit cell along wires \cite{PhysRevLett.113.086101}. The wire segment with a length of odd multiples of $a_{0}$ has an energetically unfavorable dislocation \cite{PhysRevLett.100.146103}. In this work, we focus only on the defect-free even-$a_{0}$-length wires. As shown in Fig. \ref{f1}, we can sample wire segments with varying lengths from 4$a_{0}$ to 28$a_{0}$ for a systematic length-dependent study. The very long segment of 28$a_{0}$ is important to clarify the atomic structure of a pristine wire, which is not perturbed by adatoms. A detailed topography for the central unit cells of the 28$a_{0}$ wire is perfectly matched with the simulation based on the recently proposed structural model \cite{PhysRevLett.113.086101,Tobe}. One important message is that the characteristic $\times$2 reconstruction is maintained well away from Si adatoms. This finding contradicts with the previous structure model where the $\times$2 structure is induced only by the doping from Si adatoms \cite{PhysRevB.80.155409}. This also casts strong doubts on the observation of the local doping effect of adatoms in the previous STS result \cite{PhysRevLett.109.066801}, which is discussed in detail below. 

\begin{figure}[!tb] 
\centering
\includegraphics[width=8.6cm]{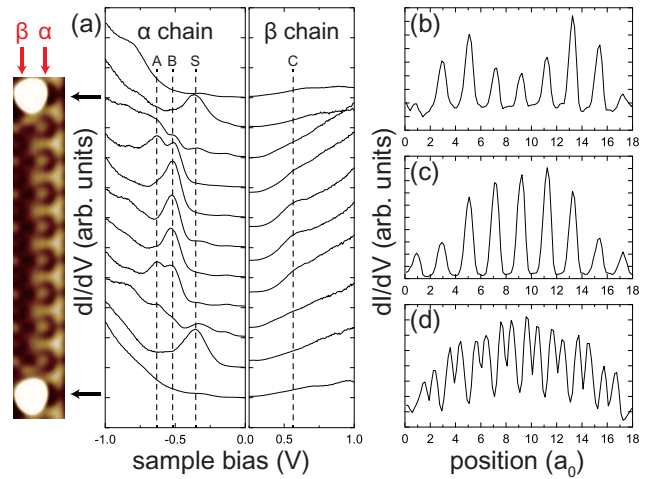}
\caption{\label{f2} (a) Spatially resolved $dI/dV$ curves of a 18$a_{0}$ wire. The curved are acquired for uniformly spaced positions along the $\alpha$ (for filled states) and the $\beta$ (for empty states) chains specified in Fig. \ref{f1} with the top and bottom curves for the adatom locations (black arrows). Four prominent peaks $A$, $B$, $C$ and $S$ are specified on the curves. $dI/dV$ spatial profiles measured for the biases (b) -0.63, (c) -0.53, and (d) +0.57 V, corresponding to the LDOS of $A$, $B$, and $C$ states, respectively.
}
\end{figure}

Electronic structures of wire segments with different lengths are investigated by the spatially resolved differential conductivity ($dI/dV$). We define two representative parts of a wire, $\alpha$ along the single 1$a_{0}$ Au chain and $\beta$ in the middle of triple 2$a_{0}$ Au chains, which exhibit distinctive spectra (see Figs. \ref{f1} and \ref{f2}). The $dI/dV$ curves on these chains change significantly not only for different lengths but also for different locations along a wire segment. We show the $dI/dV$ curves of a 18$a_{0}$ wire representatively in Fig. \ref{f2}. The characteristic electronic states of the $\alpha$ ($\beta$) chain appear noticeably in the filled (empty) state, which disappear on adatom locations (the top and bottom spectra). Four distinct spectral features, the local density of states (LDOS) maxima, are identified at -0.63, -0.53, +0.57, and -0.35 V sample biases and labeled as $A$, $B$, $C$, and $S$ states, respectively. While the state $A$ appears remarkably around 1/4 and 3/4 positions of the wire length, $B$ and $C$ states are dominant at the middle of the segment. In contrast, the $S$ state appears only in the vicinity of adatoms and is, thus, considered as an edge state of the segment or the adatom induced state, which is out of the present interest. These observations clearly evidence the strong electronic perturbation by adatoms and the strong spatial dependence in the electronic structure within a wire segment. We observe the consistent trend for whole wire lengths investigated.

The spatial variations of $A$, $B$, and $C$ states are more systematically shown in $dI/dV$ profiles for corresponding sample biases [Figs. \ref{f2}(b)$-$\ref{f2}(d)]. These data exhibit the LDOS variations of $A$, $B$, and $C$ states within the wire segment. A 2$a_{0}$-period rapid oscillation is observed commonly for all states, which reflect the $\times2$ periodicity of the lattice. The additional slowly varying envelopes, on the other hand, exhibit the electronic structure modulation by the extrinsic factor, that is, the edge adatoms. We observe two types of envelopes. That of state $A$ exhibits three nodes at the ends and the middle and two antinodes at the 1/4 and the 3/4 positions. Those for $B$ and $C$ have two nodes at the end and one antinode at the middle, being qualitatively different from the former. These envelopes are the fingerprint of QWS's of the longest ($n$=1) and the second longest ($n$=2) electron wavelengths. 

\begin{figure}[!tb] 
\centering
\includegraphics[width=8.6cm]{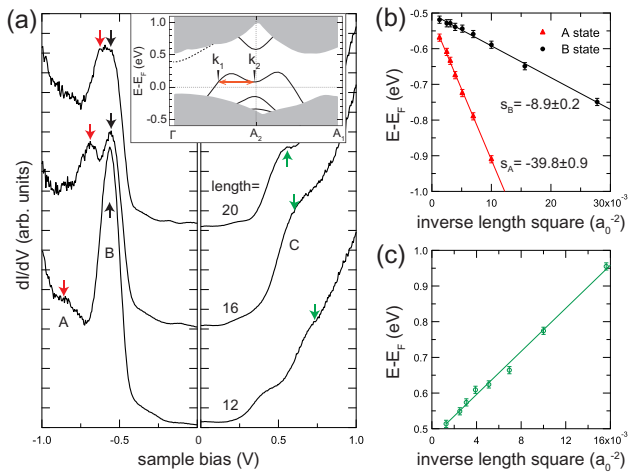}
\caption{\label{f3} (a) $dI/dV$ curves of selected wire segments (the length is given in $a_{0}$ unit) with three QWS's; $A$ (red arrows), $B$ (black), and $C$ (green) states. Filled state (empty state) curves are recorded from $\alpha$ ($\beta$) chain. Curves are vertically offset for comparison. The energies measured for states (b) $A$ and $B$ and (c) $C$ are plotted as a function of inverse-length-square of wire segments. Solid lines indicate the linear fits with the slopes for $A$ ($s_{A}$) and $B$ ($s_{B}$) are -39.8$\pm$0.9 and -8.9$\pm$0.2 (arb. units), respectively. Inset: the calculated surface state band dispersions for the adatom-free Si(111)5$\times$2-Au surface \cite{PhysRevLett.113.086101}. $\left ( \boldsymbol{k}_{2}-\boldsymbol{k}_{1} \right )/2$ is a scattering wavevector (orange arrow) \cite{hoffman2003search} near the Fermi level, which is approximately $2\pi/{8a_{0}}$. The dashed band has less weights in the topmost layer. 
}
\end{figure}

Further clarification of the QWS nature is accomplished by the extensive length dependence study. In Fig. \ref{f3}(a), we show the $dI/dV$ curves of wire segments with lengths of 12$a_{0}$, 16$a_{0}$, and 20$a_{0}$. One can clearly see the length-dependent energy shifts for at least $A$ and $C$, which approach to the Fermi level as the length increases. There also exist small energy shifts for $B$ as shown in Fig. \ref{f3}(b). This length dependence can be analyzed quantitatively with the simple 1D quantum box model. If we consider that two edge adatoms of a wire segment function as confinement potentials, 1D electrons within a segment would be quantized. The $n$-th QWS energy is given by
\begin{equation}\label{eqnenrg}
E_{n}=E_{F}+E_{0}\pm\frac{h^{2}}{8m_{e}}\left( \frac{n}{L} \right )^{2},
\end{equation}
where $E_{F}$ is the Fermi level, $E_{0}$ the band edge energy, $h$ the Planck constant, $m_{e}$ the effective electron mass, and $L$ the wire length. The plus and minus signs stand for the electron- and hole-like bands, respectively. Different band origins cause the electron wavelength and, in turn, the QWS energy changes in opposite ways. Figures \ref{f3}(b) and \ref{f3}(c) compare the length dependent energies of states $A$, $B$, and $C$. They are unambiguously depends on inverse squares of lengths (1/$L^2$) as expected from the quantum box model. Moreover, the slope for $A$ is approximately four times larger than that for $B$, which is also expected in the model for $n$=2 and $n$=1 QWS's. This is consistent with spatial envelopes of these states discussed above. These analyses confirm solidly the 1D electron quantization within wire segments. 

Based on this model, the difference in the sign of the slopes ($A$/$B$ and $C$) would indicate the hole and electron band nature of the corresponding bands, respectively. This is consistent with the band dispersions calculated shown in the inset of Fig. \ref{f3}(a) \cite{PhysRevLett.113.086101}. As expected qualitatively, the filled (empty) state bands are hole (electron) like. For a more quantitative comparison, the electronic state (band) energies for an infinite wire are estimated from a linear extrapolation of the experimental data. The fits yield $E_{0}$=-0.51 eV for the hole-like band ($A$ and $B$ merging) and +0.47 eV for the electron-like band ($C$). This result is reasonably consistent with the band calculation; in very good agreement with the band edge energy of especially the empty surface state. In filled states, there is a quantitative difference in the extrapolated and calculated energies, which comes partly from the ambiguity due to the presence of multiple surface state bands. Nevertheless, we can confidently conclude that the wire segment determined by two edge Si adatoms exhibits the 1D electron quantization. That is, the Au-Si metallic atomic wires are segmented by Si adatoms into an ensemble of quantum rods.

\begin{figure}[!tb] 
\centering
\includegraphics[width=8.6cm]{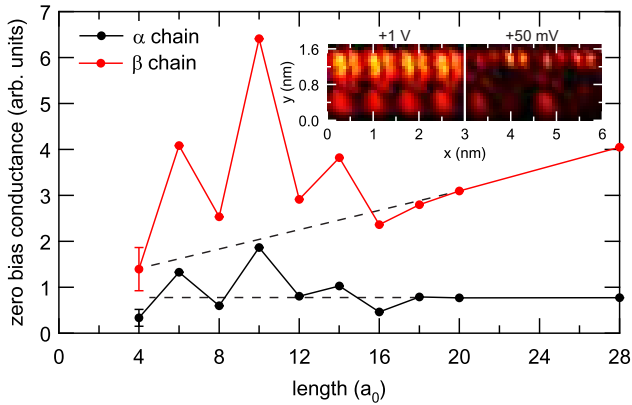}
\caption{\label{f4} Length dependent zero bias conductances measured for $\alpha$ and $\beta$ chains. All the conductance values are extracted from the average over corresponding chains to wipe out fine spatial modulations. Dotted lines display the extrapolation of saturated curves at long chains. Inset: $dI/dV$ maps measured for four central 5$\times$2 unit cells of the 28$a_{0}$ wire segment at +1 (left) and +50 mV (right) biases.
}
\end{figure}

The electron quantization commonly causes the significant oscillatory change in various physical and chemical quantities \cite{PhysRevB.27.1991,PhysRevLett.80.5381,chiang2000photoemission,sanchez1999gold}. All those quantum oscillations basically come from the DOS oscillation at the Fermi level due to the matching of the quantum well width with the multiple of half the Fermi wavelength. In the present case, the wire has one metallic band whose principal Fermi wavelength (2$\boldsymbol{k}_{1}$) is close to 4$a_{0}$. This yields only a trivial 2$a_{0}$ quantum oscillation since we have only wires with a length of even multiples of $a_{0}$. However, due to the characteristic dispersion of this metallic band, a 8$a_{0}$ Fermi scattering wavelength also emerges, corresponding to difference between $\boldsymbol{k}_{1}$ and $\boldsymbol{k}_{2}$ [the orange arrow in the inset of Fig. \ref{f3}(a)] \cite{hoffman2003search}. Note that the exact Fermi level of the present surface would deviate from that of the calculation for an adatom-free surface. This leads to the quantum oscillation of a 4$a_{0}$ periodicity in the DOS near the Fermi energy. Figure \ref{f4} shows the length dependence of the zero bias conductance measured on $\alpha$ and $\beta$ chains.  Both start with the lowest conductance at the shortest 4$a_{0}$ chain and alternatively increase and decrease at 2$a_{0}$ interval until they reaches a length of 18$a_{0}$. In our quantization scenario, the 4$a_{0}$ periodicity must have a direct correlation with the spatial electronic modulation at the Fermi level. In the inset of Fig. \ref{f4}, the $dI/dV$ maps measured for four central unit cells of the 28$a_{0}$ segments are shown. While the well-known 2$a_{0}$ periodicity of the lattice is viewed at a high bias of +1 V, a clear 4$a_{0}$ periodicity is observed near the Fermi level of +50 mV. One can, thus, easily expect that various physical and chemical properties of individual wire segments would exhibit 4$a_{0}$ quantum oscillations too, which can be exploited in further studies. 

In contradiction with the present conclusion, the previous STM/S work interpreted the length-dependent energy shift of wire segments as the confined doping effect of adatoms \cite{PhysRevLett.109.066801}. This interpretation provides the length dependence of 1/$L$ instead 1/$L^2$. This difference is thought to come from the limited energy resolution and data sets of the previous work, which were not sufficient to differentiate the 1/$L$ and  1/$L^2$ dependence and to resolve the $A$ and $B$ QWS. All the spectroscopic information and the spatial dependence of each spectral features presented here are not compatible with the local doping picture. This conclusion is fully supported by the recent structure model and the band structure calculation with and without adatoms \cite{PhysRevLett.113.086101}. The gradual change of the band dispersions observed previously as a function of the adatom density \cite{PhysRevLett.100.126801} has to be reinterpreted as to represent the gradual evolution from the band dispersions of the Si(111)5$\times$2-Au pristine surface to those of the 5$\times$4-Au structure saturated fully with Si adatoms \cite{PhysRevLett.113.086101}. This naturally explains the mysterious non-rigid shifts of the bands observed previously, which could not easily be reconciled with the doping picture \cite{PhysRevLett.100.126801}.   

In summary, Si adatoms on the Si(111)5$\times$2-Au surface confine electrons on metallic atomic wires, resulting in the 1D electron quantization. This quantization is well supported by the energy and spatial variation of the relevant electronic states and in good consistency with the band structure calculated based on a very recent structure model. The confinement also leads to a 4$a_{0}$-period oscillation of the LDOS at the Fermi level, which are expected to bring about various quantum oscillation in physical and chemical properties of these wire segments. With the established possibility of the Si adatom manipulation, the present result promises exciting future development of exploiting quantum rods based on self-organized atomic wires.

This work was supported by IBS-R014-D1. We would like to acknowledge fruitful discussions with S. G. Kwon and M. H. Kang.

\bibliographystyle{apsrev4-1} 
\bibliography{Qconfinement_bibtex} 

\end{document}